\documentclass[review]{elsarticle}
\usepackage{amsmath, amssymb, amsthm, enumitem}
\usepackage{placeins}

\pdfoutput=1
\usepackage{color}

\usepackage{IEEEtrantools}
\newcommand{\multicol}[2]{\IEEEeqnarraymulticol{#1}{#2}}
\usepackage{booktabs}

\usepackage{natbib} 
\usepackage{hyperref}

\usepackage{caption}
\usepackage{subcaption}
\usepackage{accents}

\numberwithin{equation}{section}

\begin{document}
\begin{frontmatter}

\title{Disease Momentum: Estimating the Reproduction Number in the Presence of 
Superspreading}

\author[1]{Kory D. Johnson\corref{cor1}} 
\ead{kory.johnson@wu.ac.at}
\author[2]{Mathias Beiglb\"ock\fnref{fn1}}
\ead{mathias.beiglboeck@univie.ac.at}
\author[2]{Manuel Eder\fnref{fn1}}
\ead{manuel.eder@univie.ac.at}
\author[2]{Annemarie Grass\fnref{fn1}} 
\ead{annemarie.grass@univie.ac.at}
\author[2]{Joachim Hermisson\fnref{fn1}} 
\ead{joachim.hermisson@univie.ac.at}
\author[2]{Gudmund  Pammer\fnref{fn1}}
\ead{gudmund.pammer@univie.ac.at}
\author[2]{Jitka Polechov\'a\fnref{fn1}}  
\ead{jitka.polechova@univie.ac.at}
\author[2]{Daniel Toneian\fnref{fn1}} 
\ead{daniel.toneian@univie.ac.at}
\author[2]{Benjamin W\"olfl\fnref{fn1}}
\ead{benjamin.woelfl@univie.ac.at}

\affiliation[1]{
organization={Vienna University of Economics and Business},
addressline={Welthandelsplatz 1},
city={Vienna},
postcode={1020},
country={Austria}}
\affiliation[2]{
organization={University of Vienna},
addressline={Oskar-Morgenstern-Platz 1},
city={Vienna},
postcode={1090},
country={Austria}}

\cortext[cor1]{Corresponding author}
\fntext[fn1]{Contributed equally}


\begin{abstract} 

A primary quantity of interest in the study of infectious diseases is the 
average number of new infections that an infected person produces. This 
so-called reproduction number has significant implications for the disease 
progression. There has been increasing literature suggesting that 
superspreading, the significant variability in number of new infections caused 
by individuals, plays an important role in the spread of SARS-CoV-2. In this 
paper, we consider the effect that such superspreading has on the estimation of 
the reproduction number and subsequent estimates of future cases. Accordingly, 
we employ a simple extension to models currently used in the literature to 
estimate the reproduction number and present a case-study of the progression of 
COVID-19 in Austria. Our models demonstrate that the estimation uncertainty of 
the reproduction number increases with superspreading and that this improves the 
performance of prediction intervals. Of independent interest is the derivation 
of a transparent formula that connects the extent of superspreading to the width 
of credible intervals for the reproduction number. This serves as a valuable 
heuristic for understanding the uncertainty surrounding diseases with 
superspreading.

\end{abstract}

\end{frontmatter}

\maketitle


\section{Introduction}

The reproduction number, $R_t \equiv R$, 
gives the average number of new infections caused by a single infected person 
throughout the infectious period. In contrast to the basic reproduction number 
$R_0$, which describes the reproduction of the virus in a na\"ive, unmitigated 
population, $R$ (sometimes called the \emph{effective} reproduction number) 
varies through time as the epidemic develops and the opportunities for 
transmission change due to, for example, behavioral response, seasonality, and 
changes in the relative size of the susceptible population. In every 
population, some individuals will cause considerably more infections 
than others - a phenomenon known as \emph{superspreading}.  It can be quantified 
using a framework provided by \citet{lloyd2005superspreading}. In this paper, we 
extend the model of \citet{cori2013new} to include the phenomenon of 
superspreading. Our goal is to better quantify the uncertainty inherent in this 
type of estimate of $R$, \emph{not} to derive a more accurate estimate.


Ultimately we are interested in the estimation of $R$ and specifically the 
question whether, given current case numbers, we can claim with statistical 
guarantees that $R \leq 1$ or $R > 1$. Given the growing body of evidence about 
the existence and importance of superspreaders \citep{Adam2020clustering, 
Liu2020secondary}, we incorporate this feature into our models. We observe two 
important effects: first, it becomes increasingly difficult to accurately 
estimate the reproduction number $R$ in the presence of superspreading; second, 
models with superspreading produce prediction intervals for new cases that have 
improved coverage compared to those without superpreading. Both of these are 
demonstrated in our Austrian case-study in Section \ref{sec:austria}. In 
particular, it becomes infeasible even in early May to support the claim that 
$R<1$ using our methods. This is a critical period of time as it 
coincides with the removal of lockdown restrictions in Austria.


In particular,  the width of a credible interval for $R$ should decrease as a 
function of total number of cases used during estimation and increase with the 
extent of superspreading. Let $S$ be the set of days used to estimate $R$ in the 
nowcasting framework presented in Section \ref{sec:nowcasting} and assume that 
the (average) reproduction number does not change over time. One would then 
expect that a $(1-\alpha)\%$ credible interval to have width approximately 
equal to 
\begin{IEEEeqnarray}{rCl} 
  \frac{2z_{1-\alpha/2}}{\sqrt{k \sum_{s \in S} I_s}}, 
\end{IEEEeqnarray} 
where $z_{1-\alpha/2}$ is the $(1-\alpha/2)$ quantile of the standard normal 
distribution and for values of dispersion parameter $k$ much smaller than 1, 
which corresponds to scenarios with high superspreading. We derive this exact 
functional form in a simplified model introduced in Section 
\ref{sec:generation}.

\subsection{Nowcasting}
\label{sec:nowcasting}


The goal of nowcasting is to get accurate estimates of the current state of an 
epidemic. Given that our observed infections are random observations from an 
underlying process, our goal is to understand the parameters of that process, 
particularly with respect to the reproduction number. In addition, we define a 
time-varying parameter we call the ``momentum'' of an epidemic, which is a 
\emph{random} realization of population infectiousness at a time-point which 
accounts for superspreading. This is introduced formally in Section 
\ref{sec:model}.

Benchmark methods for estimating the reproduction number $R$ include those of 
\citet{cori2013new} and \citet{WallingaT04}. The method of \citet{cori2013new} 
provides near real-time estimation of $R$ and is implemented in the R software 
package `EpiEstim'. An improvement of this framework is given in 
\citet{thompson2019improved} which accounts for variability in the generation 
interval (defined below). A substantial extension of the EpiEstim-package 
(`EpiNow') was developed by a group of researchers at the London School of 
Hygiene and Tropical Medicine \citep{abbott2020estimating}. The method of 
\citet{WallingaT04} provides an alternate estimate for historical values of $R$. 
Contrary to the methods discussed in this paper, it requires observations from 
both before and after the time point at which an estimate for $R$ is desired. 
An important overview of other estimation methods and challenges due to 
COVID-19 is given in \citet{gostic2020practical} and a comparative analysis of 
statistical methods to estimate $R$ is given in \citet{o2020comparative}. If the 
epidemic is at an early stage, the reproduction number $R$ and the rate of 
exponential growth are  connected by the Euler-Lotka 
equation \citep{wallinga2007generation, ma2020estimating}.

As we follow the framework of \citet{cori2013new}, we briefly describe their 
basic model. Let $I_0$ be the number of initial 
infections and $I_1, I_2, \ldots$ be the number of new infections on days 
$1,2,\ldots$. By $(w_n)_{n\geq 1}$ we denote the \emph{generation interval 
distribution}. If $J_m$ denotes the number of people infected by a specific 
person on the $m$-th day after this person got infected, then we have for $m\in 
\mathbb{N}$ 
$$w_m=\frac{\mathbb{E}[J_m]}{\sum_{l=1}^\infty \mathbb{E}[J_l]}.$$
We assume that a newly infected individual does not cause secondary 
cases on the same day, corresponding to $w_0=0$. The generation interval can be 
interpreted as the infectiousness profile of infected persons.

The basic model of \citet{cori2013new} assumes that the stochastic process of 
total new infections on day $t$, $(I_t)_{t\in \mathbb{N}}$, satisfies
\begin{align}
\label{Cori13model} 
I_t \sim \text{Poisson}\left(R_t \sum_{m=1}^{t}I_{t-m}w_m\right),
\end{align}
for a sequence of numbers $R_t,\,t\in\mathbb{N}$. 
In practice it is often assumed that the generation interval distribution is 
given as a  Gamma distribution that has been discretized in such a way that 
$w_m=0$ for all $m$ larger than some cut-off number $\nu$ 
\citep{gostic2020practical}. As a result, the sum in \eqref{Cori13model} will 
only have $\nu \in \mathbb{N}$ summands, and to make assertions about $I_t$ we 
only have to consider the case numbers $I_{t-\nu}, \ldots, I_{t-1}$. As $\nu$ is 
a parameter that can vary between diseases, this term is kept and used 
throughout our model description in Section \ref{sec:model}. 

When estimating the time-varying reproduction number, \citet{cori2013new} assume 
that the reproduction number has stayed constant over a window of $\tau$ days. 
In this case, for $s \in (t-\tau+1,\ldots,t)$, equation \eqref{Cori13model} 
simplifies to
\begin{align}
\label{Cori13model2} 
I_s \sim \text{Poisson}\left(R \sum_{m=1}^{\nu}I_{s-m}w_m\right).
\end{align}
In order to treat $R$ as fixed in the above expression, it is necessary to only 
explicitly model a subset of time points, lest $R$ be assumed constant over all 
time points.

Note that the reproduction number in the sense of \eqref{Cori13model2} does not 
denote the number of people that actually have been infected by a given 
individual, but rather  describes what one would expect in an ``average'' 
evolution of the epidemic. Furthermore, while $R=R_t$ is assumed to be constant 
over the window of width $\tau$, as this window moves through time the method 
produces \emph{estimates} of $R$ that slowly vary over time.

\subsection{Heterogeneity in Reproduction Numbers}
\label{sec:heterogeneity}

The motivation for our hierarchical Bayesian approach follows the framework of 
superspreading provided in \citet{lloyd2005superspreading}. Even if the 
reproduction number $R$ is constant over a small window of time, it might 
vary between individuals. We consider the reproduction number of a 
specific person with index $x$ to be drawn randomly as 
\begin{IEEEeqnarray}{rCl}
r_x \sim \text{Gamma}(k,\, \text{rate} = k/R). \label{eqn:r_x}
\end{IEEEeqnarray}
This distribution has mean $R$ and variance $R^2/k$. Note that the above gamma 
distribution will also be referred to as having dispersion parameter $k$. The 
degenerate case $k=\infty$  corresponds to the deterministic case where $r_x= R$ 
for all individuals and leads to the model in \eqref{Cori13model2}. Given 
$r_x=r$, this person causes Poisson($r$) new infections. 
If one integrates out the Poisson parameter $r$, one is left with the 
unconditional number of descendants which follows a negative binomial 
distribution with mean $R$ and variance $R+R^2 / k$. This negative binomial 
model is further analyzed in Section \ref{sec:generation}.

A basic 
extension of \eqref{Cori13model2} that follows the concept of random individual 
reproduction numbers in the sense of \citet{lloyd2005superspreading} is to 
assign, on day $t$, the individual reproduction numbers $r_1^t, \ldots, 
r_{I_t}^t$ to the $I_t$ individuals that got infected on this day. This 
leads to the recursion 
\begin{align}\label{Kory20}
I_t \sim \text{Poisson}\left(\sum_{m=1}^\nu w_m \sum_{x=1}^{I_{t-m}} 
r_x^{t-m}\right),
\end{align}
where the individual reproduction numbers $r_x^{m}$ are drawn i.i.d.\ 
according to \eqref{eqn:r_x}. Note that for the degenerate case $k=\infty$, 
\eqref{Kory20} recovers \eqref{Cori13model2}. This forms the foundation of the 
model explained in detail in Section \ref{sec:model}.

The theme of the present paper is close to that of 
\citet{donnat2020modeling}, in which heterogeneity in $R$ between \emph{groups} 
is explicitly modeled. While the high-level descriptions of these models sound 
nearly identical, those models are relevantly different than ours. In 
particular, \citet{donnat2020modeling} are interested in estimating 
group-specific or time-varying reproduction numbers for different geographical 
regions and age groups. On one hand, with sufficient group-specific data, this 
provides tools of a much broader scope than we present here; on the other 
hand, it is assumed that within-group variability is negligibly small. Instead, 
we focus on aggregate data from a \emph{single} geographical region but do 
\emph{not} assume that individual variability is negligible. Rather,  this is 
precisely the variability we are interested in modeling. Furthermore, our 
critiques of the estimability of the reproduction number transfers to their 
setting as well: if within-group variability exists, group-specific reproduction 
numbers are more difficult to estimate than previously acknowledged.

\section{Methods}

This section introduces two methods. First, the ``momentum'' model formulates 
the estimation problem as a Bayesian Poisson regression. Second, the 
``generation'' model is a simplification which provides a fast approximation to 
the momentum model as well as an explicit formula for dependence of credible 
interval width on $k$. Both are of interest beyond COVID modeling and aim 
to address different goals: precise estimation (momentum) and valuable speed and 
heuristics (generation).


\subsection{The ``Momentum'' Model}
\label{sec:model}

%

As mentioned in the introduction, we identify an unobserved random variable 
which we term the ``momentum'' of the epidemic. This follows from a simple 
notational change in \eqref{Kory20} according to the observation that a sum of 
i.i.d. Gamma random variables is also Gamma distributed with the same 
dispersion parameter. We rewrite \eqref{Kory20} as
\begin{align}\label{Kory20.2}
I_t \sim \text{Poisson}\left(\sum_{m=1}^\nu w_m \theta_{t-m}\right),
\end{align}
where
\begin{IEEEeqnarray}{rCcCl}
  \label{Kory20.2b}
 \theta_{t} & = & \sum_{x=1}^{I_t} r_x^t
 & \sim & \text{Gamma}(I_tk,\, \text{rate} = k/R).
\end{IEEEeqnarray}
The terms $(\theta_t)_{t\geq 0}$ are collectively referred to as the 
``momentum'' of the disease. They will be treated as a set of nuisance 
parameters of the offspring distribution, as our primary interest lies in 
estimating the reproduction number $R$. In our Bayesian framework introduced 
below, $R$ is a hyperparameter of the prior distribution for $(\theta_t)_{t\geq 
0}$.
Equation \eqref{Kory20.2} describes the distribution
of $I_t$ conditioned on its whole past, i.e., $I_{s}, \theta_{s}$, $s < t$.
Analogously, equation \eqref{Kory20.2b} describes $\theta_s$ given its history.
The difference in what we understand as the relative past originates from
$\theta_t$ being conceptually determined ``after'' $I_t$.

For increased clarity of the form of the model and the estimation methods 
required, we recast our model as a Bayesian Poisson regression using vector 
notation. This is made painfully explicit by using an arrow as in $\vec{I}$ for 
vectors. Following \citet{cori2013new}, we estimate $R$ by explicitly modeling a 
set of $\tau$ days over which we assume $R$ to be constant. We specify the 
regression function for each observation in this estimation window. To condense 
notation, we use $[l]$, for $l\in\mathbb{N}$, to be the vector $(1,2,\ldots,l)$. 
Similarly, $[l,\,m]$ for $l,\,m\in\mathbb{N}$ is shorthand for the vector 
$(l,l+1,\ldots,m)$, i.e., $[l]=[1,l]$. This notation will primarily be used for 
vector indices. Furthermore, the indices of our vectors increase in time. As 
such, our generation interval truncated to $\nu$ days can be condensely written 
as $\vec{w}_{[\nu]} = (w_1, \ldots, w_\nu)$. Similarly, the $\tau$ observations 
we model are given by $\vec{I}_{[t-\tau+1,\,t]} = (I_{t-\tau+1}, \ldots, I_t)$.

As a regression model for $\vec{I}_{[t-\tau+1,\,t]}$, equation \eqref{Kory20.2} 
can be written as
\begin{IEEEeqnarray*}{rCl}
 \vec{I}_{[t-\tau+1,t]} & \sim & \text{Poisson}(W 
\vec{\theta}_{[t-\nu-\tau+1,t-1]})\quad \text{where} 
\IEEEyesnumber\label{eq:reg}\\
W &=& 
\begin{pmatrix}
w_\nu & w_{\nu-1} & \ldots & w_1 & 0 & 0 & \cdots & 0 \\
0 & w_\nu & w_{\nu-1} & \ldots & w_1 & 0 & \cdots & 0 \\
\vdots  & \ddots & \ddots & \ddots & \ddots & \ddots & \cdots &\vdots  \\
0 & \cdots & 0 & w_\nu & w_{\nu-1} & \ldots & w_1 & 0\\ 
0 & \cdots & 0 & 0 & w_\nu & w_{\nu-1} & \ldots & w_1
\end{pmatrix}\\ 
\end{IEEEeqnarray*}
In the above expression, we have a fixed covariate matrix $W$ which is a 
function of the generation interval $w_{[\nu]}$. The momentum parameters 
$\vec{\theta}_{[t-\nu+\tau-1,t-1]}$ are seen to be the regression parameters to 
be estimated. Note that the expressions in the previous display suppress the 
notation for conditioning on all observations before time $t-\tau+1$. 
Furthermore, given $\vec{\theta}_{[t-1]}$, $I_t$ is independent of 
$\vec{I}_{[t-1]}$. 

We place a prior distribution on $\vec\theta$ 
which depends on $R$ as in equation \eqref{Kory20.2b}, as well as a hyperprior 
on $R$ to account for the previously identified uncertainty in the distribution 
of $R$ as reported in \citet{abbott2020estimating}. As we have parameterized 
the gamma prior on $\theta_t$ to have mean $I_tR$, the conjugate hyperprior for 
$R$ is the inverse-gamma distribution. This is transparent in the posterior 
distribution given by equation \eqref{eq:post} below. Hence we use an 
inverse-gamma hyperprior on $R$, where these hyperparameters are set to 
match the results of \citet{abbott2020estimating}. As such, we assume that $R$ 
has mean 2.6 and standard deviation 2, yielding shape parameter $3.69$ and rate 
parameter $6.994$:
\begin{IEEEeqnarray*}{rCl}
R & \sim & \text{Inv-Gamma(3.69, rate = 6.994)}.
\end{IEEEeqnarray*}
An a priori distribution for $R$ is itself uncertain and one 
could theoretically place additional hyperpriors on the parameters of this 
inverse-gamma distribution. That being said, the change would increase
computational complexity while introducing hyper-hyperparameters that would 
be difficult to estimate. Hence, this proposal distribution for $R$ is treated 
as fixed.

This regression formulation is important as it highlights the 
latent variables $\vec{\theta}$ that are required to fully determine the 
generative model. It also focuses attention on which observations are 
conditioned upon and which are treated as random, i.e., the $\tau$ observations 
to which we fit the model are treated as random. This is relevant as more 
than $\tau$ nuisance parameters are present, namely $\nu+\tau-1$. Observe that 
the earliest data point is $I_{t-\tau+1}$, which itself requires a history of 
$\nu$ momentum values of $\vec{\theta}$ to determine.

While we also think of individual reproduction numbers as changing over time due 
to factors such as changes in social restrictions, the assumption of constant 
$R$ over a period renders this moot. Likewise, we set $k$ to be a constant for
the results presented in Section \ref{sec:austria}, as $k$ is best estimated 
with 
contact tracing data instead of case count data. We set $k=0.072$, in line with 
the results of \citet{Lax+20}, which estimated the extent of superspreading for 
COVID-19 from Indian data. This is also within the range of parameter values 
identified in \citet{Endo+20}.

Alternatively, it is possible to consider an independently estimated 
distribution for $k$. To estimate the momentum model with random $k$, one can 
merely draw $k$ from a proposal distribution and estimate the momentum model 
with this fixed value. This process is repeated for many sampled values of $k$,
and the posterior samples for $R$ and $I_t$ from all $k$ are combined. This 
follows the same methodology as \citet{thompson2019improved}, where the 
generation interval was estimated with a separate data set before fitting 
model \eqref{Cori13model2} without superspreading. Brief results for this case 
are presented in \ref{sec:validation} as none of the results change 
significantly. The joint estimation of $k$ and $R$ within the momentum model 
appears infeasible as $k$ is the dispersion parameter of the nuisance parameter 
distribution. This makes learning about $k$ using this data highly challenging.



A full derivation of the posterior distribution of the pair 
$R,\vec{\theta}_{[t]}$ given $\vec{I}_{[t]}$ is given in \ref{sec:derivation}. 
We obtain as posterior
\begin{IEEEeqnarray*}{rCl}
\multicol{3}{l}{p(R,\vec{\theta}_{[t - \tau -\nu + 1, \, t-1]}|\vec{I}_{[t - 
\tau - \nu + 1, \, t]})}\\
  & \propto & p(\vec{I}_{[t-\tau+1,\,t]}, \vec{\theta}_{[t - \tau + 1,\, 
t-1]}|\vec{\theta}_{[t - \tau - \nu + 1, \, t - \tau]},\vec{I}_{[t - \tau - \nu 
+ 1, \, t-\tau]},R)
  p(\vec{\theta}_{[t - \tau]},R|\vec{I}_{[t-\tau]})\\
  & \propto & \left(\prod_{s = t - \tau + 1}^t \Big(\sum_{m < s} w_{s - m} 
\theta_m\Big)^{I_{s}} 
e^{- \sum_{m < s}  w_{s - m} \theta_m  } \right)\\
  & & \cdot \left(\prod_{s = t - \tau + 1}^{t-1} 
\frac{k^{I_sk}}{\Gamma(I_sk)R^{I_sk}}
\theta_s^{I_sk-1}e^{-\frac{k}{R} \theta_s} \right)
\left(\prod_{s=t-\nu-\tau+1}^{t - \tau} 
\frac{k^{I_sk}}{\Gamma(I_sk)R^{I_sk}}
      \theta_s^{I_sk-1}e^{-\frac{k}{R} \theta_s}\right)\\
  & & \cdot 
      \left(R^{-3.69-1} e^{-6.994/R}\right),\IEEEyesnumber\label{eq:post}.
\end{IEEEeqnarray*}
The first line of \eqref{eq:post} specifies the distribution of the observations 
given all other parameters, and the third line gives the inverse-gamma prior for 
$R$. The second line describes the distribution of $\vec\theta$, and we have 
explicitly partitioned the indices into two sets. The values $\theta_s$ in the 
first index set $[t-\tau+1,t-1]$ require no special discussion as they depend 
on values $I_s$ which are being explicitly modeled. The values of $\theta_s$ 
in the second index set $[t-\nu-\tau+1,t-\tau]$, however, treat the 
corresponding $I_s$ values as \emph{fixed} and \emph{constant}. This is done 
so that 
we do not need to specify further nuisance parameters before time $t - 
\tau - \nu + 1$. Doing so would create an infinite recursion in historical 
observations, requiring us to treat $R_t$ as fixed for all $t$. Hence we need 
not only a prior for $R$, but also for $\vec{\theta}_{[t - \tau - \nu +1, \, t 
- \tau]}$. More details are provided in \ref{sec:derivation}.

In order to condense notation for summations in exponents, let $S$ be the index 
set for the second product; i.e., $S = 
\{t-\nu-\tau+1,t-\nu-\tau+2,\ldots,t-1\}$. The additional shorthand below drops 
``$s\in$'' from $s\in S$. With this notation, the posterior distribution of $R$ 
given $\vec{\theta}$ and $\vec{I}$ is
$$p(R|\vec{\theta}_{[t-1]},\vec{I}_{[t]}) \propto R^{-k 
\sum_{S}I_s-3.69-1} 
  e^{(-k\sum_{S}\theta_s - 6.994)R^{-1}},$$
which is Inv-Gamma($k \sum_{S}I_s+3.69$, $k\sum_{S}\theta_s + 6.994$). 
A perhaps counter-intuitive observation is that the posterior distribution of 
$R$ does not depend on the generation interval $\vec w_{[\nu]}$. This is the 
result of conditioning on $\vec{\theta}$ versus integrating it out as done in 
\citet{lloyd2005superspreading}. In our case, it is infeasible to integrate out 
$\vec\theta$ as the dependence is too complex. If we truly 
know population infectiousness, i.e., the epidemic momentum at all points in 
time, then $\vec w_{[\nu]}$ is irrelevant for estimating $R$, because $\vec 
w_{[\nu]}$ just determines \emph{how we learn} about $\vec{\theta}$ via 
\eqref{eq:reg}. More concretely, there are no terms in \eqref{eq:post} that 
include all of $R$, $\vec{\theta}$, and $w_{[\nu]}$.

The posterior expectation and variance of $R$ are
\begin{IEEEeqnarray*}{rCl't}
\mathbb{E}[R|\vec{\theta},\vec{I}] & = & \frac{k\sum_{S}\theta_s + 6.994}{k 
\sum_{S}I_s+3.69-1} & and\\
\text{Var}[R|\vec{\theta},\vec{I}] & = & \frac{(k\sum_{S}\theta_s + 
6.994)^2}{(k 
\sum_{S}I_s+3.69-1)^2(k \sum_{S}I_s+3.69-2)}. 
\end{IEEEeqnarray*}

The denominator of the variance picks up an additional $k$ term, making 
credible intervals wider when $k$ is small. The dependence on $\vec\theta$ is 
difficult to remove in this general setting. Section \ref{sec:generation} 
considers a simpler setting in which $\vec\theta$ can be integrated out in 
order to derive a transparent function for credible interval width.

%

To estimate this model, we  alternate between a Gibbs-step to sample $R$ and a 
Metropolis-Hastings step to sample $\vec{\theta}$. As 
$\mathbb{E}[\theta_s|I_s,R] = I_sR$, we can initialize reasonable starting 
values for $\vec{\theta}$ using various values of $R$ such that we require 
little burn-in. We find total chain length to be the more important tuning 
parameter for valid prediction and credible intervals. In all models presented 
in this paper, we set $\nu=\tau=13$ to make valid comparisons with results from 
the EpiEstim framework \citep{cori2013new}. We set $\vec w_{[\nu]}$ to be a 
discretized gamma distribution with mean 4.46 and standard deviation 2.63 per 
the results of \citet{agesInterval} for Austria, which are similar to values 
determined elsewhere \citep{Knight20, Ganyani2020}. Inference is conducted using 
the $10^6$ samples that remain after a burn-in of 1,000 and thinning by 5. 

While the majority of the model validation and supporting graphs is relegated 
to 
\ref{sec:validation}, we address here the particular concern that we have 25 
nuisance parameters in $\vec\theta$ for modeling 13 observations. Our simulation 
evidence indicates that all nuisance parameters are well-estimated, even those 
far in the past: coverage of $\vec\theta$ by credible intervals in simulated 
data is nearly exact. Furthermore, we see approximate coverage when predicting 
new cases in Section \ref{sec:austria}. As such, we do not believe that we are 
over-fitting the data with a larger number of nuisance parameters. This is in 
part due to the role of the prior distribution for $\theta_s$. For example, the 
first nuisance parameter $\theta_{t-\nu-\tau+1}$ only appears in a single 
observation term in the posterior \eqref{eq:post}: the distribution of 
$I_{t-\tau+1}$. Similarly, $\theta_{t-\nu-\tau+2}$ only appears in two, etc. The 
prior therefore plays a larger role in determining the values of these 
parameters.

\subsection{Generation Model}
\label{sec:generation}

In order to directly relate the dispersion parameter $k$ to the width of the 
credible interval and to provide a fast approximation to the momentum model, we 
consider the trivial generation interval in which an infected person is only 
infectious for a single day. For real data, this assumption is obviously 
inaccurate. Therefore, we switch to modeling infections per generation instead 
of infections per day. While we model generations spanning 
multiple days, we estimate and forecast cases for conventional days.



When the generation interval $w$ is of this form, $\vec{w}_{[1]} = (1)$, the 
model is purely Markovian and the data follow a Galton-Watson process. Recall 
that a Poisson$(\lambda)$-distributed random variable $Y$, where $\lambda$ is 
distributed according to Gamma$(\alpha,\beta)$, follows a 
negative binomial distribution \citep{lloyd2005superspreading}:
\begin{equation}
    \label{eq:NB}
    Y \sim NB\left(\alpha,\frac{1}{1 + \beta}\right), \quad p(Y) = 
\frac{\Gamma(Y + \alpha)}{Y!\Gamma(\alpha)}\left(\frac{\beta}{ 1 + 
\beta}\right)^\alpha \left(\frac{1}{1 + \beta}\right)^Y.
\end{equation}
Applying \eqref{eq:NB} and $\vec{w}_{[1]} = (1)$ to the momentum model 
\eqref{Kory20.2} yields the 
following distribution for the infections $I_t$:
\begin{IEEEeqnarray}{rCl}
    \label{eq:infections trivial 1}
    I_t | \vec{I}_{[t-1]},R,k & \sim & NB \left( k I_{t-1}, \frac{R}{R + k} 
\right),
\\  \label{eq:infections trivial 2}
    p(I_t| \vec{I}_{[t-1]}, R, k) & = &\frac{\Gamma(I_t + 
kI_{t-1})}{I_t!\Gamma(kI_{t-1})} \left(\frac{k}{R + k}\right)^{k I_{t-1}} 
\left(\frac{R}{R+k}\right)^{I_t}.
\end{IEEEeqnarray}

In \ref{sec:meta-derivation}, we reparameterize this model in terms of 
$\frac{R}{R+k}$ in order to place a suitable prior which mimics that of the 
momentum model. After transforming the resulting posterior back to a 
distribution for $R$ and using standard normal approximation techniques 
\citep{gelmanBDA04}, we derive a normal approximation of the posterior of
\begin{IEEEeqnarray*}{rCl}
p(R|\vec{I}_{[t]},k) & \approx & N\left(\frac{k(\alpha - 1)}{\beta + 
1},\frac{k^2(\alpha + \beta)(\alpha -1)}{(\beta + 1)^3}  \right). 
\end{IEEEeqnarray*} 
where
\begin{IEEEeqnarray*}{rCl't'rCl}
\alpha & = & 98.82 + \sum_{ s = t - \tau + 1}^t I_s & and &
\beta & = & 3.74 + k \sum_{s = t - \tau}^{t-1} I_s.
\end{IEEEeqnarray*}

We are interested in the setting in which $R\approx 1$ and $\beta \approx 
k\cdot\alpha$. Note 
that $\sum_{s = t - \tau + 1}^t I_s$ and $k \sum_{s = t - \tau}^{t - 1} I_s$ 
are of this approximate ratio: the terms in these two sums almost 
entirely overlap. Furthermore, while the hyperparameters (98.82 and 3.74) are 
of moderate size, they also approximately satisfy the desired ratio. This 
yields the following simplification of the variance of the normal approximation:
\[
    \frac{k^2(\alpha + \beta)(\alpha - 1)}{(\beta + 1)^3} \approx 
\frac{k^2\alpha^2(k+1)}{k^3\alpha^3} = \frac{k+1}{k\alpha} \approx \frac{1}{k 
\sum_{s = t - \tau + 1}^t I_s}.
\]
Hence, the approximate length of a credible interval for $R$ behaves like
\[  \frac{2z_{1-\alpha/2}}{\sqrt{k \sum_{s = t - \tau + 1}^tI_s}}.    \]

It is clear that the assumption $\nu=1$ is highly unrealistic for COVID-19 and 
most other diseases. In order to bridge this gap, we estimate the model for 
non-overlapping generations instead of conventional days.
The length of a generation is set equal to the mean of the generation interval, 
i.e.,
\[
    D_{g} := \sum_{t = 0}^{\nu} t \omega_t.
\]
Given the modeling assumptions we have made for COVID-19, a 
generation comprises approximately 4.87 conventional days. The first 4.87 days 
after infection also accounts for 64\% of the assumed infectiousness given by 
the generation interval. This helps explain why partitioning the 
data into generations produces reasonable results. When a model is defined over 
generations, setting $\nu=1$ is equivalent to assuming that someone is equally 
infectious over $D_{g}$ days. The negative binomial model estimated using 
generations is approximately equivalent to the momentum model estimated using 
conventional days. 


In order to account for non-integer-valued generations, consider 
$D_{g} = 
\lfloor D_{g} \rfloor + D_{frac}$, where $D_{frac} \in [0,1)$. For 
simplicity, we assume that new infections are uniformly distributed during the 
day so that we may use standard data with records of new daily cases. In 
order to not confuse subscripts indexing days and generations, times in the 
generation model will be indicated by $\tilde t$ instead of $t$. Lastly, as we 
are interested in using the most recent data, we care about matching the right 
endpoint of our time series. As such, we compute the generations 
\emph{backwards} from a reference day $t$. 

Let day $t$ be the maximal day in our data set. We define the 
corresponding generation incidence, $\tilde I_{\tilde t}$, to be
\begin{IEEEeqnarray*}{rCl}
  \tilde I_{\tilde t} & = & \sum_{s = 0} ^{\lfloor D_{g} \rfloor - 1} 
I_{t-s} + D_{frac} \cdot I_{t - \lfloor D_{g} \rfloor}.
\end{IEEEeqnarray*}
This is merely the sum over $\lfloor D_{g} \rfloor$ full days, and a 
proportion of the remaining day. Infections for previous generations then sum 
similarly over the historical data such that the generations form a partition 
of 
days in our data set. 

As before, some mathematical details are moved to 
\ref{sec:meta-derivation}. With simple notational changes, however, we derive a 
model for generations which looks functionally identical to 
\eqref{eq:infections 
trivial 1}, i.e.,
\[
\tilde I_{\tilde t} | R, \tilde I_{\tilde t-1} \sim NB\left( \tilde 
I_{\tilde t-1}k, \frac{R}{k+R}\right).
\]
This formula can then be used to forecast the cumulative incidence over several 
generations as described in \ref{sec:meta-derivation}. This yields a simple, 
closed form approximation of the momentum model without resorting to costly 
Bayesian computation methods.

\section{Results}
\label{sec:austria}

This section focuses on understanding the evolution of the reproduction number 
in Austria between April 1 and October 31, 2020. As the momentum model 
effectively needs $\tau+\nu$ observations to be fit, this is approximately as 
early as estimates can be provided for Austria. Our goals are three-fold: to 
demonstrate the increase in estimated variability of $R$ due to superspreading, 
to provide valid prediction intervals for new cases, and to compare to similar 
models without superspreading. Some results will be shown for Croatia and 
Czechia as well to help establish the validity of our method, but the focus in 
on Austrian data. Other supporting graphs for Croatia and Czechia are given in 
\ref{sec:results2}.

An important component of estimating the reproduction number on a given date is 
to account for the delay distribution between date of infection and 
date of confirmation as discussed in \citet{gostic2020practical}. If a delay of 
length $d$ occurs between infection and confirmation, then an infection 
observed at time $t$ actually occurred on day $t-d$. In this case, we have a 
``true infection history'' that is distinct from the reported case numbers. In 
reality, the delay $d$ is random. \citet{abbott2020estimating} estimate and 
sample true potential infection histories given observed case numbers by 
sampling possible delays $d$. As our primary goal is to understand the 
uncertainty in estimating $R$ as opposed to providing best in class predictions 
of $R$ for a given date, we ignore this complication. This allows us to take as 
model input the historical 7-day moving average of reported cases and to compare 
methods with simple, transparent input. As a result, however, we are not 
attempting to predict the number of true infections on a given date. Instead, we 
are predicting the number of reported or confirmed cases on this date. In order 
to highlight this, axes are explicitly labeled with ``Reported Cases'' and 
``Confirmation Date''.

\begin{figure}
\centerline{\includegraphics[width=1\textwidth]{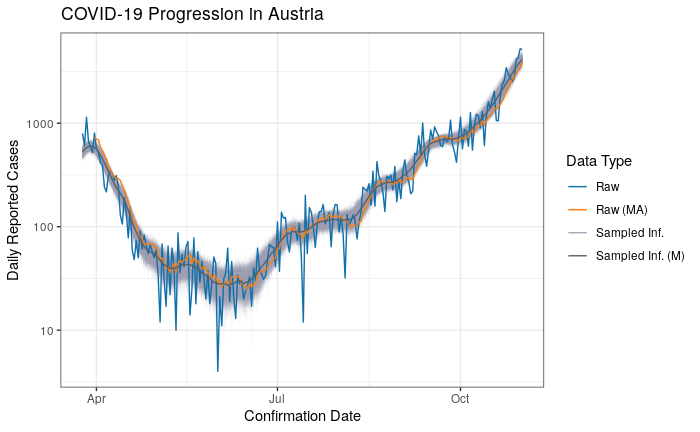}}
\caption{Summary of new cases of COVID-19 in Austria: raw infection data (Raw), 
the 7-day moving average of Raw (Raw (MA)), each sampled infection history 
(Sampled Inf.), and the daily median of the sampled infection histories (Sampled 
Inf.\ (M)).}
\label{fig:austria-history}
\end{figure}

Data on the progression of COVID-19 in Austria is shown in Figure 
\ref{fig:austria-history}. This graph includes curves for the raw infection 
data as reported by the European Center for Disease Prevention and Control 
(Raw), the 7-day moving average of Raw (Raw (MA)), each sampled infection 
history (Sampled Inf.), and the daily median of the sampled infection histories 
(Sampled Inf.\ (M)). Observe that the boundary of the ``band'' created by the 
sampled infection histories is not smooth, as it is created from 1,000 distinct 
faded lines. Note that using sampled infection histories effectively shifts the 
time series backward in time. In order for the infection histories to 
approximately match the reported case numbers, we have aligned them in time.

As mentioned in Section \ref{sec:model}, we sample one million total samples of 
$R$ and the momentum vector $\vec{\theta}$. To forecast future cases, we use an 
individual sample of parameters and run the momentum model for a specified 
period of time. Our graphs show results for the average number of new cases over 
the following week. As such, they are on the scale of daily reported cases. 
There is no additional smoothing of the raw data or predictions. As our input is 
the 7-day moving average, our prediction is the 7-day-ahead forecast of this 
moving average. 

In all of the following graphs, we plot predictions and intervals from three 
models: the momentum model with $k=0.072$, the generation model of Section 
\ref{sec:generation} with $k=0.072$, and the EpiEstim model of 
\citet{cori2013new}. 
As mentioned previously and visible in \ref{sec:results2}, treating $k$ as 
random within a relevant region does not alter our results.
We label the EpiEstim model ``Epi*'', as the estimates are produced 
directly via equation \eqref{eq:epistar} below instead of using the 
EpiEstim R package.
As in \citet{cori2013new}, we fix a generation interval, as opposed 
to taking samples of a generation interval estimated from a separate data source 
as in \citet{thompson2019improved}. As a result, we are not comparing to 
the best in class model within the EpiEstim/EpiNow framework, but with a 
model of corresponding complexity to the momentum model. Other improvements to 
the modeling framework could then be built on top of the momentum model as they 
have been for the model of \citet{cori2013new}.

To estimate the model of \citet{cori2013new}, we estimate the parameters of the 
\citet{cori2013new} posterior distribution directly from the infection data:
\begin{IEEEeqnarray}{rCl}
\label{eq:epistar}
 p(R_t|I_{[t]}) & = & \text{Gamma}\left(a + \sum_{s=t-\tau+1}^t I_s,\, \text{rate} = b +
   \sum_{s=t-\tau+1}^t \sum_{m = 1}^\nu w_{m}I_{s - m}\right)
\end{IEEEeqnarray}
where $a$ and $b$ are the shape and rate parameter of the gamma prior 
distribution on $R$. We estimate this posterior distribution, draw one million 
samples for $R$, and run the corresponding data generating process 
\eqref{Cori13model} for the required number of days.

\begin{figure}
\centerline{\includegraphics[width=1\textwidth, trim=0 0 0 0, 
clip]{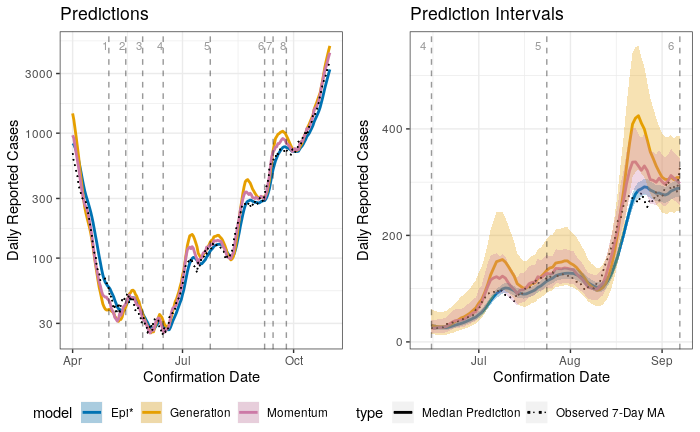}}
\caption{Predictions between April 1 and October 31, 2020, and 90\% prediction 
intervals between two significant dates: June 15 and September 7, 2020. 
Predictions and intervals are for the 7-day average of new cases in the 
following week in Austria. Relevant event dates are given as vertical, dashed 
lines and are described in Table \ref{tab:austria-dates}. The Epi* predictions 
consistently lag behind the observed values, whereas the other methods overshoot 
in the peaks due to momentum. Models with superspreading produce predictions 
intervals 2-3 times as wide as those without and achieve better coverage.}
\label{fig:austriaPI}
\end{figure}

\begin{table}
\begin{center}
\caption{Dates of important events related to COVID-19 in Austria. 
Changes which occur in large parts of the country but not uniformly are listed 
as occurring in ``some regions''.}
\label{tab:austria-dates}
\begin{tabular}{lll}
  \toprule
  Label & Date & Event\\\midrule 
  NA & 2020-03-16 & Start of general lock down \\
  1 & 2020-05-01 & Begin relaxation of movement restrictions\\ 
  2 & 2020-05-15 & Bars and restaurants can open\\ 
  3 & 2020-05-29 & Hotels and cultural sites can open\\ 
  4 & 2020-06-15 & Near complete removal of COVID restrictions\\ 
  5 & 2020-07-24 & Face masks mandatory in essential businesses\\ 
  6 & 2020-09-07 & Start of school year in some regions\\ 
  7 & 2020-09-14 & Face masks mandatory\\ 
  8 & 2020-09-25 & Bars and restaurants close early in some regions \\
  NA & 2020-11-03 & Start of general soft lock down \\\bottomrule
\end{tabular}
\end{center}
\end{table}

Figure \ref{fig:austriaPI}  shows the difference between models with and without 
superspreading on Austrian data. In order to show a long time period, the data 
must be plotted on a logarithmic scale such that the low cases in the summer 
months are visible. As this distorts the plotting of prediction intervals in the 
same graph, the comparison of prediction intervals is given separately by 
focusing attention on the summer months between the effective end of COVID 
restrictions and the start of the school year. 

\begin{figure}
\centerline{\includegraphics[width=1\textwidth, trim=0 0 0 0, 
clip]{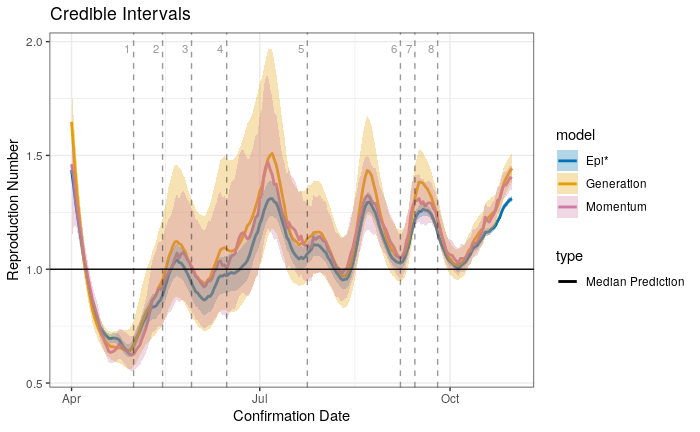}}
\caption{Credible intervals for R in Austria. The momentum and generation model 
predictions are consistently slightly higher than those of Epi*. They also 
produce credible intervals that are 2-3 times as wide. Relevant event dates are 
given as vertical, dashed lines and are described in Table 
\ref{tab:austria-dates}. Observe that $R$ becomes indistinguishable from 1 using 
our models around the time when lockdown restrictions begin to be removed.}
\label{fig:austriaCI}
\end{figure}

For reference, we marked the dates of important changes in COVID-19 restrictions 
in Austria as vertical, dashed lines. A complete list is available at
\href{https://regiowiki.at/wiki/Chronologie_der_Corona-Krise_in_Österreich}
{https://regiowiki.at} (in German).
The events are described in Table \ref{tab:austria-dates}. When 
comparing the events to both reported cases and the estimated reproduction 
number in Figure \ref{fig:austriaCI}, it is necessary to keep the delay 
distribution in mind; i.e., the effect of an intervention will not be visible in 
confirmed cases and thereby the estimated reproduction number for roughly two 
weeks \citep{abbott2020estimating}. Prior to the removal of any lockdown 
restrictions, reported case numbers were decaying exponentially. This is visible 
as a linear decrease given the logarithmic scaling of the y-axis. The slope of 
this line changed substantially around the time that Austria began to reopen in 
May and June. From approximately July through the end of October, case numbers 
fluctuate between growing exponentially and brief periods of relative stability. 
These fluctuations are not modeled and reflect both noise as well as features 
which we do not include in our analysis, e.g., common holiday periods, changes 
in testing, etc. Throughout this period, some restrictions are brought back 
into effect without apparent substantial impact. Lockdown measures were 
reinstated at the end of the plotted window of time. 

While all of the prediction curves track the observed cases, there are subtle 
but significant differences in behavior. If one looks closely, one can see that 
the Epi* model predictions lag behind the observed 7-day moving average: it 
fails to accurately estimate the rapid changes in case numbers. On the other 
hand, the momentum and generation model predictions ``overshoot'' the peaks in 
the 
time series. As the name suggests, there appears to be excess ``momentum'' in 
the process around these change points, and the model anticipates cases to 
continue rising as in the previous days.

The various models produce prediction intervals with drastically different 
widths. Most notably, the intervals for the momentum model with $k=0.072$ are 
much wider than those of Epi*. The generation variant of this model produces 
intervals which are wider still. The momentum intervals are, on average, 
approximately three times as wide as those of Epi*. While the generation model 
provides a computationally cheap and fast estimate, it is clear that it suffers 
relative to the momentum model in terms of interval length. The ratio between 
the prediction interval lengths visible during the summer months is 
approximately the same throughout the entire prediction period.

To assess the validity of the prediction intervals, Table \ref{tab:coverage} 
shows, for each method, the proportion of true weekly new cases that fall within 
the prediction intervals over the prediction period. Coverage is shown for the 
50\% and 90\% prediction intervals for the raw infection data. When cases are 
steadily increasing (or decreasing) prediction intervals become narrower, and 
when the behavior changes they become considerably wider. The prediction 
intervals of the momentum model cover the true values during periods of growth, 
while those of Epi* often fail to do so over the entire growth period. Clearly 
coverage is still not exact, and all models perform worse on the Czech data 
(see \ref{sec:results2}). It is still notable that the momentum models provide 
approximate coverage in these cases even with the inherent messiness of the 
COVID-19 case data. For example, Czechia had a much higher test positivity rate 
than Austria and Croatia during the majority of the prediction period, which is 
ignored in our model.

\begin{table}
\begin{center}
\caption{Coverage of the 50\% and 90\% prediction intervals (PI) for 
7-day-ahead predictions of the 7-day moving average. Models with superspreading 
improve coverage significantly over that of Epi*.}
\label{tab:coverage}
\begin{tabular}{llcc}
  \toprule
  Country & Model & Coverage, 50\% PI & Coverage, 90\% PI\\\midrule
  Austria & Momentum, k = 0.072 & 0.46 & 0.79 \\ 
   & Generation, k = 0.072 & 0.47 & 0.73 \\
   & Epi*, k $\rightarrow \infty$ & 0.16 & 0.38 \\\midrule
  Croatia & Momentum, k = 0.072 & 0.48 & 0.85 \\ 
   & Generation, k = 0.072 & 0.49 & 0.77 \\
   & Epi*, k $\rightarrow \infty$ & 0.18 & 0.47 \\\midrule
  Czechia & Momentum, k = 0.072 & 0.40 & 0.69 \\
   & Generation, k = 0.072 & 0.39 & 0.66 \\
   & Epi*, k $\rightarrow \infty$ & 0.12 & 0.32 \\\bottomrule
\end{tabular}
\end{center}
\end{table}

As the reproduction number is unobserved, we are unable to compare our 
predictions within a supervised setting as we compared our model forecasts. 
Given the previous discussion though, we see that the additional variability 
provided by the momentum model is needed to provide prediction intervals with 
approximate coverage. Figure \ref{fig:austriaCI} shows the median predictions 
and 90\% credible intervals for $R$ given by the momentum, generation, and Epi* 
models.  Intervals are, in general, asymmetric, and skewed toward higher values. 
The figure clearly demonstrates that the intervals for $R$ are drastically 
different: with superspreading, intervals for $R$ are roughly 2-3 \emph{times} 
as wide as those without. This could have potentially large implications for 
policy making as we know that relatively small changes in the size of $R$ can 
lead to large differences in the number of new cases if the disease is allowed 
to progress unchecked.

Near the beginning of our estimation period and around the time when 
restrictions were being relaxed in Austria, it quickly becomes infeasible to 
claim that the reproduction number is below 1; i.e., the credible intervals 
estimated during May and June include the value 1. Beginning in 
July and August, however, we observe long periods with reproduction numbers 
significantly greater than 1, even with our comparatively wide credible 
intervals. As before, there is a delay of approximately two weeks between when 
these interventions occur and any change in reproduction number could be 
observed. Hence any discussion of dates should be interpreted loosely.

As we see a clear improvement in coverage for switching to a model with 
superspreading, it is useful to have a clearer understanding of the degree of 
heterogeneity implied by our models. To do so, we consider the posterior 
samples of $R$ from October 31, 2020. According to equation \eqref{eqn:r_x}, 
each individual has a separate reproduction number, $r_x$, given the population 
reproduction number $R$. For each posterior sample of $R$, we therefore 
draw an individual $r_x$ and secondary infections $I_x$. The Epi* models of 
\citet{cori2013new} set $r_x=R$ for all individuals. Hence, it is possible to 
compare the degree of heterogeneity by considering a Lorenz curve of the 
population of values of $r_x$ or $I_x$ \citep{Lorenz}.

\begin{figure}
\centering
  \begin{subfigure}{.49\textwidth}
   \centerline{\includegraphics[width=1\textwidth, trim=90 0 90 0, 
clip]{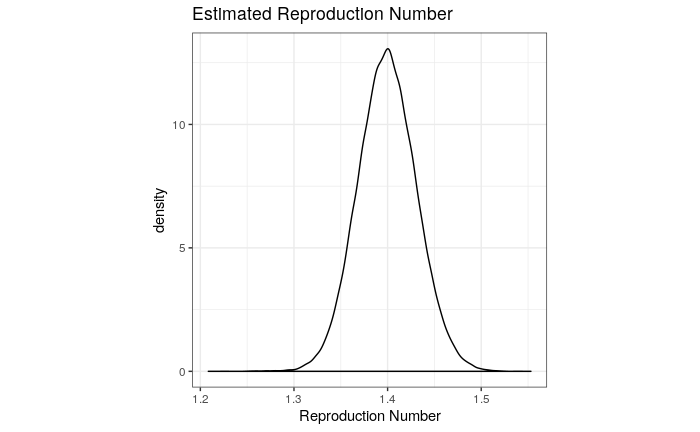}}
   \caption{Estimated distribution of $R$}
   \label{fig:R_dens}
  \end{subfigure}
  \begin{subfigure}{.49\textwidth}
   \centerline{\includegraphics[width=1\textwidth, trim=90 0 90 0, 
clip]{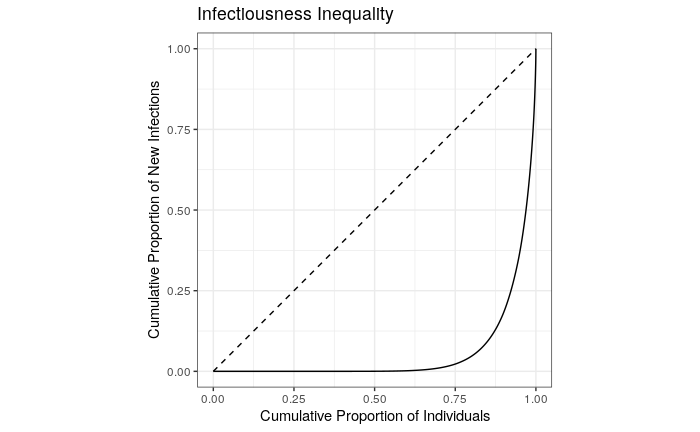}}
   \caption{Lorenz curve of $r_x$}
   \label{fig:lorenz-sub}
  \end{subfigure}
  \caption{Momentum model estimates of $R$ and individual heterogeneity for 
October 31, 2020. 10\% of individuals are expected to contribute approximately 
84.6\% of new infections. The dashed curve in (b) corresponds to a model without 
superspreading (Epi*).}
  \label{fig:lorenz}
\end{figure}

The Lorenz curve is typically used to demonstrate income inequality by 
showing the proportion of overall income or wealth held by the bottom 
x\% of the people. Here we consider this to be ``infectiousness inequality''. 
The distribution of $R$ estimated for October 31, 2020 as well as the implied 
Lorenz curve are shown in Figure \ref{fig:lorenz}. The Lorenz 
curve is a representation of the cumulative distribution function of the 
number of new expected infections. It allows us to visualize the degree of 
heterogeneity by seeing which proportion of individuals contribute to new 
infections. One can draw the Lorenz curve with $I_x$ instead of $r_x$, which 
only results in a slightly rougher image with no qualitative differences.

While the population reproduction number is moderately high, this is largely 
driven by superspreading. The momentum model implies that the top 10\% of 
individuals contribute 84.6\% of new infections, while the top 20\% contribute 
98\%. The usefulness of Figure \ref{fig:lorenz-sub} is that is shows this 
entire distribution instead of these two common quantiles. We can clearly see 
that essentially no new cases are produced by nearly 75\% of infected 
individuals.  These statistics match quite closely the observed values reported 
in \citet{Ari+2020}. The figures can also be drawn for the estimation setting in 
which $k$ is assumed to be randomly drawn from an appropriate gamma 
distribution. The resulting graphs look essentially identical. As such, treating 
$k$ as fixed at $0.072$ or fluctuating in the approximate range $[.04,\,.2]$ 
makes little difference in the infectiousness inequality implied by the momentum 
model.

\section{Conclusion}

In this paper, we provide a simple extension of the \citet{cori2013new} model to 
account for superspreading. While we explicitly use this to model the 
COVID-19 pandemic, the methods are easily adaptable to other diseases where 
superspreading is present. This ``momentum'' model incorporates unobserved 
random variables which drive the process of new infections. Even if case numbers 
and $R$ are relatively small, the presence of superspreaders can increase the 
momentum of the disease beyond what would be expected if all individuals have 
the same infectiousness. We observe that this appears necessary to properly 
track the steep increases or decreases in reported COVID-19 cases. The momentum 
model produces credible intervals and posterior predictive intervals that are 
approximately 2-3 times as wide as those that neglect
superspreading. We find that these wider intervals significantly improve the 
coverage of the prediction intervals. The heterogeneity in infectiousness 
implied by the momentum model is extremely high: 10\% of individuals 
contribute approximately 84.6\% of new infections.

As Bayesian models are time and resource intensive to estimate, we also derive a 
simplified model in which infected individuals are only infectious for a single 
day. In order to improve the fit to real data, we partition disease incidence 
into generations, each of which spans multiple days. The length of each 
generation corresponds to the generation time of the disease, and within this 
period an infected person is assumed to be equally infectious. This yields two 
main benefits. First, estimation of $R$ and predictions of new cases are 
immediately available through an explicit approximation of the posterior 
distribution of $R$. Second, this model allows us to derive a simple equation to 
relate the width of credible intervals to the degree of superspreading. Hence, 
we have rigorous analysis which supports the heuristic that the approximate 
length of a credible interval for $R$ behaves like
\[\frac{2z_{1-\alpha/2}}{\sqrt{k \sum_{s = t - \tau + 1}^tI_s}},\]
where $z_{1-\alpha/2}$ is the $(1-\alpha/2)$ quantile of the standard normal 
distribution and for values of dispersion parameter $k$ much smaller than 1, 
which corresponds to scenarios with high superspreading. The model assumes that 
$R$ has been constant for the preceding $\tau$ days.


\appendix

\section{Likelihood Derivation}
\label{sec:derivation}

This appendix derives the posterior distribution of $R$ and 
$\vec{\theta}_{[t-\tau-\nu+1, t-1]}$ given the relevant observable past, i.e., 
$\vec{I}_{[t-\tau-\nu+1, t]}$. We briefly restate some basic properties and 
definitions of our model.

Let $w_i$ denote the expected proportion of future infections caused by an 
infected person which occur on day $i$ after infection. Let $\nu$ denote the 
length of infectiousness, i.e., $w_{\nu + k} = 0 $ for all $k>0$. Lastly, 
$\tau$ denotes the number of days over which we assume $R$ is constant.

Our distributional assumptions are as follows:
\begin{align*}
    I_t | \vec \theta_{[0,t-1]},\vec I_{[0,t]}, R & \sim \text{Poisson} \left( 
\sum_{s=1}^{\nu} \omega_s 
    \theta_{t-s}\right), \text{ i.e., }
    \\      p( I_t |\vec \theta_{[0,t-1]}, \vec I_{[0,t]}, R) &= 
\frac{1}{I_t!} \left( \sum_{s=1}^{\nu} 
    \omega_s \theta_{t-s} \right)^{I_t} e^{ -  \sum_{s=1}^{\nu} \omega_s 
        \theta_{t-s}}; \text{ and}
    \\ \theta_s | R, \vec I_{[0,s]}, \vec \theta_{[0,s-1]} & \sim 
\text{Gamma}\left( I_s k, \text{rate} = \frac{k}{R}\right), \text{ i.e., 
    }
    \\      p(\theta_s | R, \vec I_{[0,s]}, \vec \theta_{[0,s-1]}) &= 
\frac{\left(\frac{k}{R}\right)^{I_s 
            k}}{\Gamma(I_s k)} \theta_s^{I_s k -1} e^{- \frac{\theta_s k}{R}
    }.
\end{align*}

We want to calculate the joint distribution:

\begin{IEEEeqnarray*}{rCl}
    \multicol{3}{l}{p(\vec \theta_{[t-\tau-\nu+1, t-1]}, R | \vec 
I_{[t-\tau-\nu+1, t]})} \\
    &= &p(\vec \theta_{[t-\tau-\nu+1, t-1]}, R | \vec I_{[t-\tau-\nu+1, 
t-\tau]},  \vec I_{[t-\tau+1, t]}) \\
    & \propto &p(\vec I_{[t-\tau+1, t]}| \vec I_{[t-\tau-\nu+1, t-\tau]},\vec 
\theta_{[t-\tau-\nu+1, t-1]}, R) p(\vec \theta_{[t-\tau-\nu+1, t-1]}, R|\vec 
I_{[t-\tau-\nu+1, t-\tau]}) \\
    & = &p( \vec I_{[t-\tau+1, t]},\vec \theta_{[t-\tau-\nu+1, t-1]}, R|\vec 
I_{[t-\tau-\nu+1, t-\tau]}) \\
    & = &p( I_{t},\theta_{t-1}|\vec I_{[t-\tau-\nu+1, t-1]},\vec 
\theta_{[t-\tau-\nu+1, t-2]},R) \\
    && \cdot p( \vec I_{[t-\tau+1, t-1]},\vec 
\theta_{[t-\tau-\nu+1, t-2]}, R| \vec I_{[t-\tau-\nu+1, t-\tau]}) \\
    & =&p( I_{t}|\vec \theta_{[t-\nu, t-1]})p(\theta_{t-1}| I_{t-1},R) p(\vec 
I_{[t-\tau+1, t-1]},\vec \theta_{[t-\tau-\nu+1, t-2]}, R| \vec 
I_{[t-\tau-\nu+1, t-\tau]}).
\end{IEEEeqnarray*}

In the last step we used the conditional independence properties for $I_t$ and 
$\theta_{t-1}$, respectively. Repeating this process to separate 
$I_{[t-\tau+2, t]}$ and $\theta_{[t-\tau+1,t-1]}$ from the rest yields:

\begin{IEEEeqnarray*}{rCl}
  \multicol{3}{l}{p(\vec \theta_{[t-\tau-\nu+1, t-1]}, R |\vec 
I_{[t-\tau-\nu+1, t]})}\\
    & \propto  & \prod_{s=t-\tau +2}^{t} p( I_{s}|\vec \theta_{[s-\nu, s-1]})\\
    && \cdot\prod_{s=t-\tau+1}^{t-1}p(\theta_{s}| I_{s},R) p( I_{t-\tau+1},\vec 
\theta_{[t-\tau-\nu+1, t-\tau]}, R|\vec I_{[t-\tau-\nu+1, t-\tau]})
\end{IEEEeqnarray*}

Now, focusing on the last term, we have

\begin{IEEEeqnarray*}{rCl}
    \multicol{3}{l}{p( I_{t-\tau+1},\vec \theta_{[t-\tau-\nu+1, t-\tau]}, 
R|\vec I_{[t-\tau-\nu+1, t-\tau]})}\\
    &=&p( I_{t-\tau+1},\vec \theta_{[t-\tau-\nu+1, t-\tau]}|\vec 
I_{[t-\tau-\nu+1, t-\tau]}, R) p(  R|\vec I_{[t-\tau-\nu+1, t-\tau]}) \\
    &=&p( I_{t-\tau+1}|\vec I_{[t-\tau-\nu+1, t-\tau]},\vec 
\theta_{[t-\tau-\nu+1, t-\tau]}, R) p(\vec \theta_{[t-\tau-\nu+1, t-\tau]}| 
\vec I_{[t-\tau-\nu+1, t-\tau]}, R) \\
    && \cdot p(  R|\vec I_{[t-\tau-\nu+1, t-\tau]}) \\
    &=&p( I_{t-\tau+1}|\vec \theta_{[t-\tau-\nu+1, t-\tau]}) 
\prod_{s=t-\tau-\nu+1}^{t-\tau} p( \theta_{s}|\vec I_{[t-\tau-\nu+1, t-\tau]}, 
R) p(  R|\vec I_{[t-\tau-\nu+1, t-\tau]}).
\end{IEEEeqnarray*}

In the last equation, we used the fact that the individual $\theta_{s}$ are 
conditionally independent given the vector $\vec{I}$. At this point, the terms 
$p(\theta_{s}|\vec I_{[t-\tau-\nu+1, t-\tau]}, R)$ become problematic. 
Knowledge of the terms $I_{m}$ for $m>s$ certainly should shed some insight on 
the value of $\theta_s$; however, it is not clear how this can be 
feasibly handled. It is not possible to prevent 
the occurrence of such terms due to the hierarchical nature of this model: the 
distribution of $I_s$ requires previous $\theta$ values, which in return demand 
the inclusion of previous $I$ values ad infinitum. This problem could be avoided 
by modeling all data from the start of the epidemic, at which point we could 
confidently set all values of $I$ and $\theta$ corresponding to times prior to 
the onset of the epidemic to $0$. This, however, would require treating the 
value of R as fixed for the entire epidemic, rendering our approach irrelevant 
as this assumption is clearly false. 

As a solution, we propose putting a prior distribution on these problematic 
$\theta_s$ such that $ p( \theta_{s}|\vec I_{[t-\tau-\nu+1, t-\tau]}, R) \sim 
\Gamma (I_s k, k/R)$, essentially disregarding the additional information 
provided by future observations. Using a different prior, such as 
setting $ p ( \theta_{s}|\vec I_{[t-\tau-\nu+1, t-\tau]}, R) = 
\delta_{RI_s}$\textemdash which has the appeal of creating terms such as those 
in \citet{cori2013new}\textemdash is statistically unsound, as we would draw 
different $\theta_s$ from different types of distributions.

All this taken together yields:
\begin{IEEEeqnarray*}{rCl}
\multicol{3}{l}{p(\vec \theta_{[t-\tau-\nu+1, t-1]}, R |\vec I_{[t-\tau-\nu+1, 
t]})}\\
    & \propto & \prod_{s=t-\tau +1}^{t} p( I_{s}|\vec \theta_{[s-\nu, s-1]}) 
\prod_{s=t-\tau+1}^{t-1}p(\theta_{s}| I_{s},R) \prod_{s=t-\tau-\nu+1}^{t-\tau} 
p( \theta_{s}| I_{s}, R)\\
    && \cdot p(  R|\vec I_{[t-\tau-\nu+1, t-\tau]})
\end{IEEEeqnarray*}

Using an inverse-gamma prior on R and using the densities of the other terms as 
discussed before evaluates to the same likelihood as in the main text. 

\section{Model Validation}
\label{sec:validation}

\begin{figure}
\centering
  \begin{subfigure}{.49\textwidth}
   \centerline{\includegraphics[width=1\textwidth]{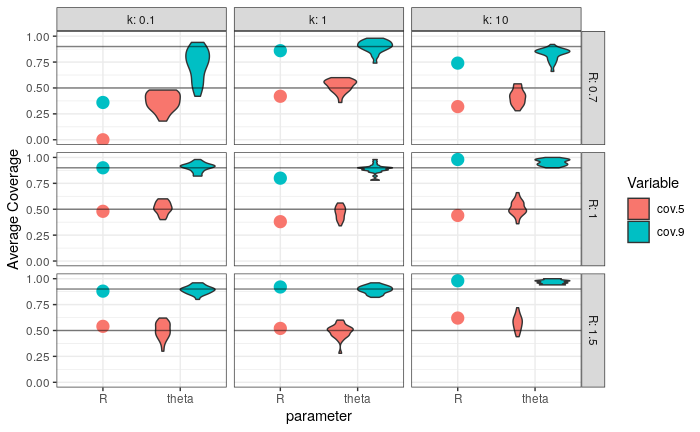}}
   \caption{}
   \label{fig:effects-1}
  \end{subfigure}
  \begin{subfigure}{.49\textwidth}
   \centerline{\includegraphics[width=1\textwidth]{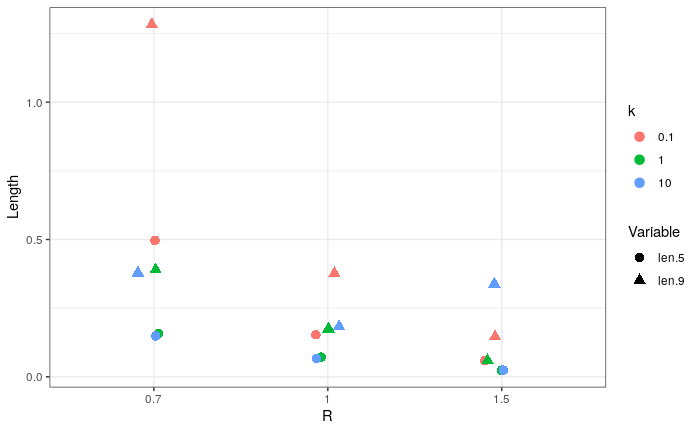}}
   \caption{}
   \label{fig:effects-2}
  \end{subfigure}
  \caption{Illustration of average credible interval coverage (cov.-) and length 
(len.-) on simulated data. As there is a single $R$ parameter but 25 elements 
of $\vec{\theta}$, the coverage of the latter are summarized via a violin plot.}
  \label{fig:cov-len}
\end{figure}

Here we summarize estimation results for simulated data in order to more 
precisely show the effect of superspreading in a setting in which true 
parameters are known. The coverage and length of intervals are shown in Figure 
\ref{fig:cov-len}. All simulations use an initial sequence of $\tau$ 
observations that have constant value 50. The momentum model is simulated for a 
further $3\tau$ days. This complete series is then used to estimate $R$ and 
$\vec\theta$. Simulations were repeated 50 times in order to asses coverage 
probabilities.

Of greatest initial import is verifying that the 90\% credible intervals for R 
indeed cover the true value with approximately nominal probability. The case 
$R=1$ is of primary importance, as it represents the bright-line between the 
epidemic growing or shrinking. That we have nearly exact coverage in this 
setting is indication that our credible intervals do not achieve coverage 
merely by being extremely wide. Furthermore, the intervals for $\vec{\theta}$ 
also cover the true values with the specified probability when $R=1$ or
$R=1.5$. With our initial sequence of cases and $R=.7$, the epidemic 
sometimes dies out, which can be missed by the model. As such, coverage 
somewhat worse in this case.

After establishing coverage, our motivation for modeling superspreading is 
verified by looking at the lengths of the credible intervals: for $k$ small, 
our intervals need to be extremely wide. In fact, the interval for $k=.1$ is 
approximately 2.5 times longer than the interval for $k=10$ for both $R=.7$ and 
$R=1$. For $R=1.5$, the estimation problem becomes relatively easy as case 
numbers grow substantially. This leads to very small credible intervals.

\begin{figure}
 \centerline{\includegraphics[width=1\textwidth]{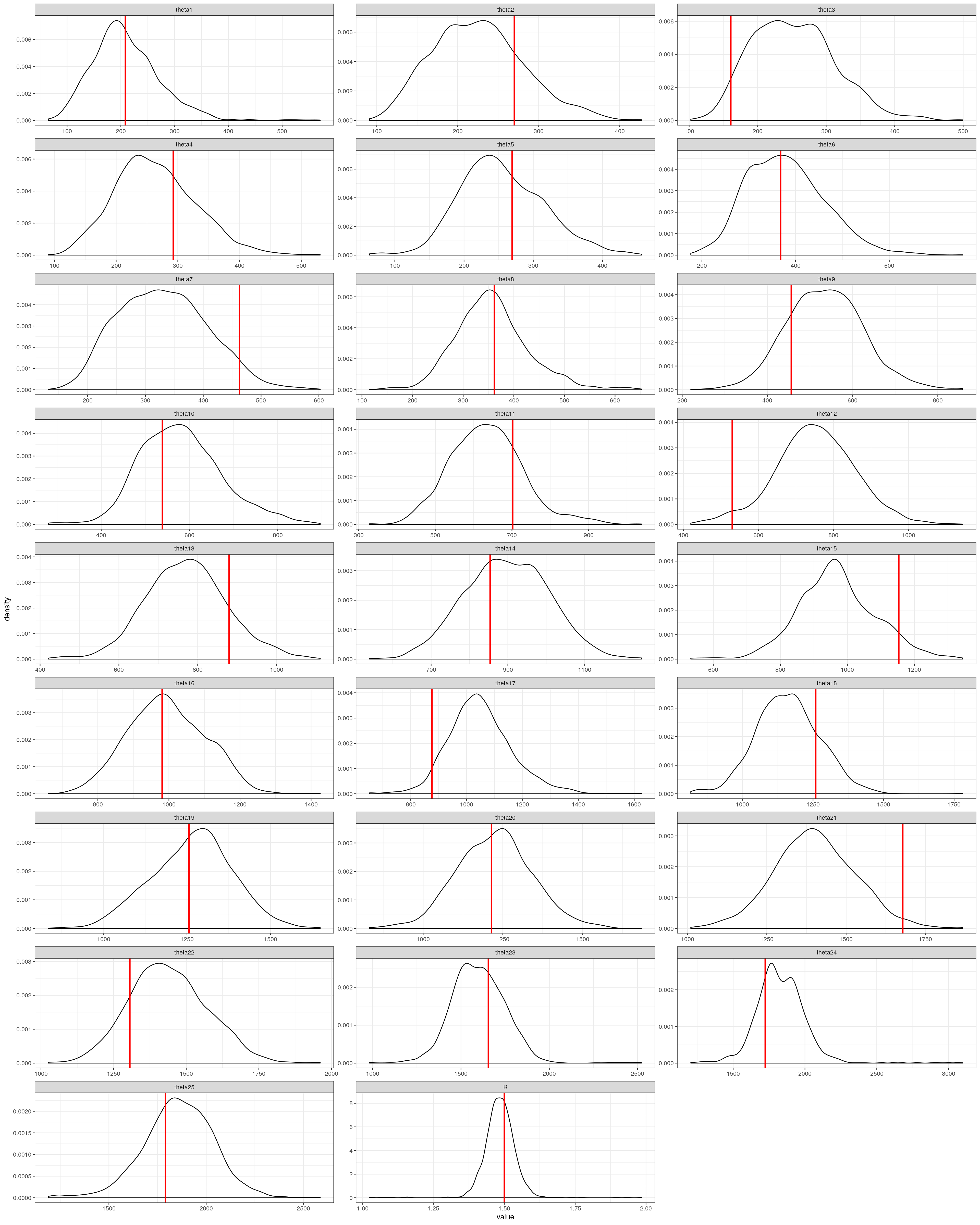}}
 \caption{Samples of MCMC draws of parameters. The vertical, red 
lines indicate true values.}
  \label{fig:par-dist}
\end{figure}

As the explicit conditional distribution of the momentum parameters 
$\vec{\theta}$ is intractable, we present a summary of the samples observed 
through the MCMC simulation in Figure \ref{fig:par-dist}. This includes all 25 
momentum parameters required when $\tau=\nu=13$ as well as $R$. As $R=1.5$ in 
this setting, one can observe that the scale increases for $\theta_s$ as $s$ 
increases. It is clear that the parameters vary widely through MCMC estimation, 
even though the are initialized at the marginal MLE: $\hat\theta_s = I_s 
\hat R$. Multiple chains are run, each with a separate initial value for $\hat 
R$. When $k$ is small, variability in $\vec{\theta}$ is large, requiring both 
tuning of the proposal distribution and long chains to be simulated in order to 
overcome high auto-correlation in the MCMC draws of $\vec{\theta}$.

\begin{figure}
 \centerline{\includegraphics[width=1\textwidth]{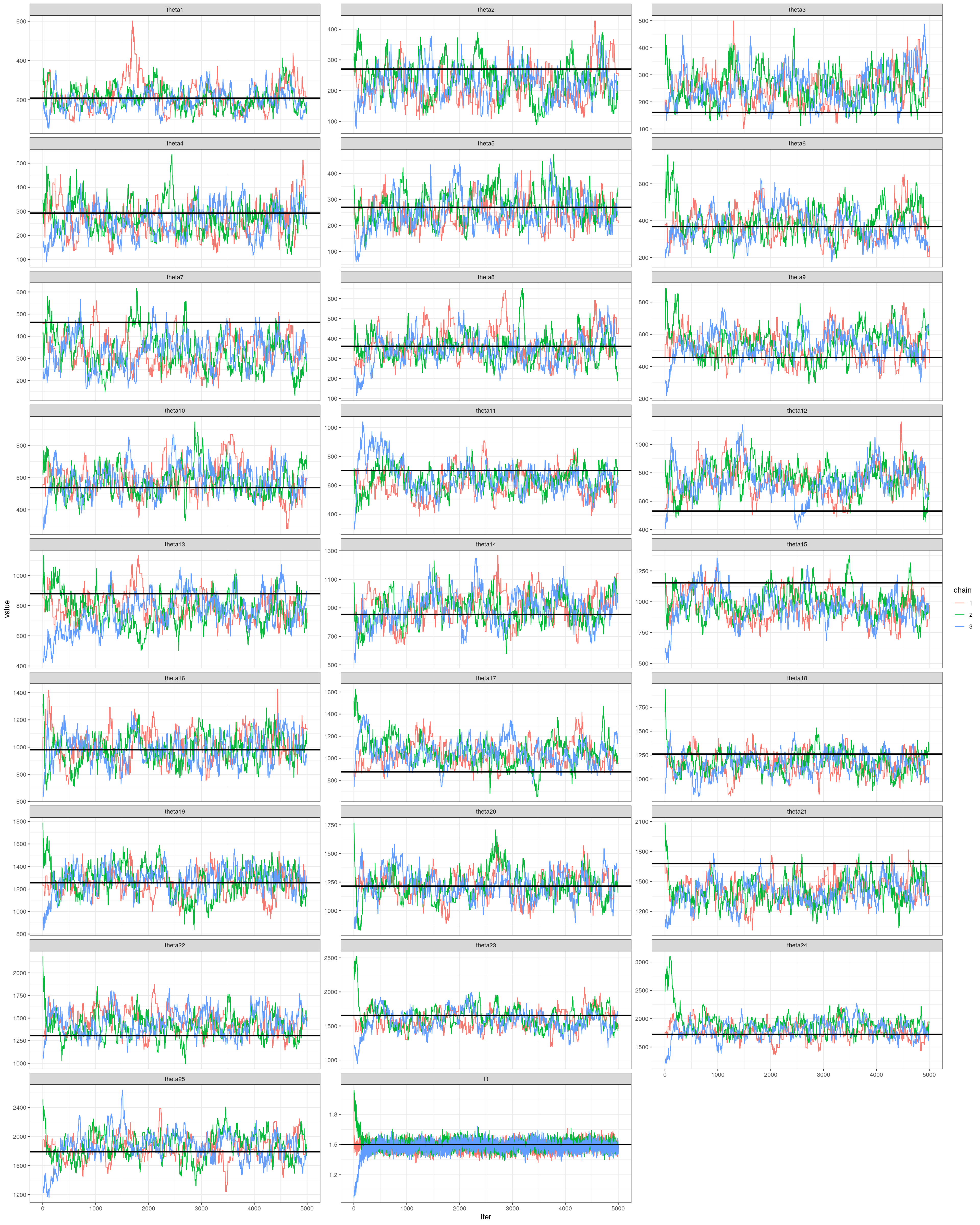}}
 \caption{Each posterior distribution is composed of samples from several 
chains. As seen above, the chains converge quickly.}
  \label{fig:chain}
\end{figure}

Figure \ref{fig:chain} shows how the individuals MCMC chains behave for each of 
the 25 momentum parameters in $\vec\theta$. Graphs for all parameters are shown 
in order to demonstrate that there is insufficient information to estimate the 
full set of parameters. One can also see how quickly the parameter estimates 
from different chains converge even when started a significantly 
different\textemdash and in some cases completely incorrect\textemdash starting 
values. Depending on the value of $k$, the variance of the proposal 
distribution for $\vec\theta$ must be set in order to allow $\theta$ to move 
slowly. If the variance is too high, then the acceptance proportion of proposed 
parameters is extremely low. This is due to the vector jumping to a nonsensical 
configuration, even if each individual $\theta_i$ is plausible in isolation.

\begin{figure}
 \centerline{\includegraphics[width=1\textwidth]{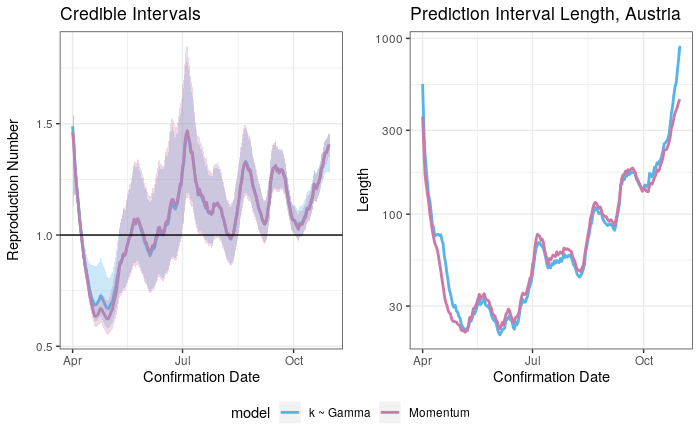}}
 \caption{Comparison of selected graphs for $k$ fixed and $k \sim Gamma()$.}
  \label{fig:kgamma}
\end{figure}

As a final model validation, we consider $k$ being drawn from a 
suitable distribution instead of being fixed. By using $k\sim$ 
Gamma(6, rate=55) we achieve approximately the same 2.5\%, 50\%, 
and 97.5\% quantiles of the distribution of $k$ given in \citet{Endo+20}. For 
reference, these are .04, .1, and .2, respectively. Figure \ref{fig:kgamma} 
shows credible intervals for R and prediction interval lengths for the momentum 
model with $k = 0.072$ and $k$ random as above. Only these summary graphs are 
shown because no differences are visible in the missing figures. The only 
notable difference in the estimation of the reproduction number occurs when 
observed cases are very low. In this region, treating $k$ as random yields 
slightly larger estimates for $R$ as well as wider confidence intervals. 
Lastly, we note that the intervals for $R$ are not as symmetric as for the 
$k$-fixed case as they are skewed slightly left. Furthermore, there is less 
heterogeneity in infectiousness. Our models estimate that 10\% of 
infected individuals contribute 81\% of new infections while 20\% contribute 
95\% of new infections.

\section{Generation Model Derivations}
\label{sec:meta-derivation}

\subsection{Normal Approximation}

This appendix derives the normal approximation to the posterior distribution 
$p(R|\vec{I}_{[t]},k)$ used in Section \ref{sec:generation}. As we can 
iteratively 
condition on previous values, the joint distribution of $\vec{I}_{[t - \tau + 1, 
t]}|\vec{I}_{t-\tau},R,k$ decomposes into a product of factors of the form 
\eqref{eq:infections trivial 2}. We have
\begin{IEEEeqnarray*}{rCl}
p(\vec{I}_{[t - \tau + 1, t]} | \vec{I}_{[t - \tau]}, R,k) 
  & = & \prod_{s = t - \tau + 1}^t p(I_s |\vec{I}_{[s-1]},R,k)\\  
  & = & \prod_{s = t - \tau + 1}^t \frac{\Gamma(I_s + 
kI_{s-1})}{I_s!\Gamma(kI_{s - 1})} \left(\frac{k}{R + k}\right)^{k I_{s-1}} 
\left(\frac{R}{R+k}\right)^{I_s}.
\end{IEEEeqnarray*}
The structure of this likelihood suggests estimating $R/(R + k)$ instead of 
$R$. 
When treating $\vec{I}_{[t - \tau]}$ and $k$ as fixed, Bayes' theorem yields 
the 
posterior distribution of $R/(R+k)$:
\begin{IEEEeqnarray*}{rCl}
\multicol{3}{l}{p\left( \frac{R}{R + k}\middle|\vec{I}_{[t]},k \right)}\\ 
& \propto & \left( 
\frac{k}{R + k} \right)^{k \sum_{s = t - \tau}^{t-1} I_s} \left(\frac{R}{R + 
k}\right)^{\sum_{s = t - \tau + 1}^t I_s} p\left( \frac{R}{R + k}\middle| I_{[t 
- \tau]}, k \right).\IEEEyesnumber \label{eq:posterior_meta}
\end{IEEEeqnarray*}

\begin{figure}
\centerline{\includegraphics[width=1\textwidth]{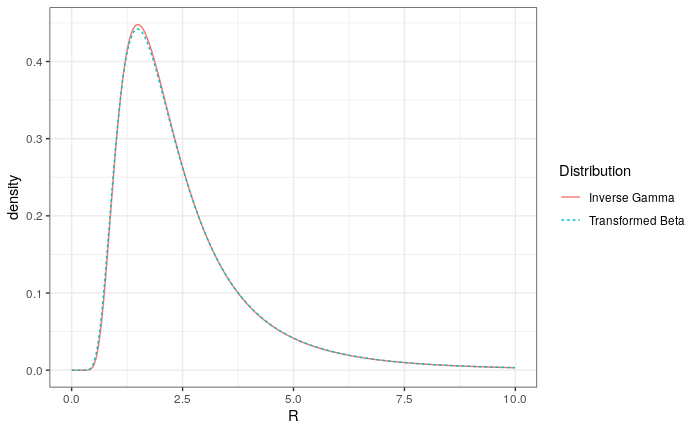}}
\caption{Comparison of priors on $R$ and $R/(R + k)$.}
\label{fig:beta}
\end{figure}

Given this functional form, it is natural to put a beta prior on $R/(R + k)$ to 
maintain conjugacy. For small $k$, as shown in Figure \ref{fig:beta}, this 
corresponds to putting an appropriate inverse-gamma prior on R, while for larger 
k this would correspond to a gamma prior. Therefore, to mimic the $R \sim 
\text{Inv-Gamma}(3.69,\, \text{rate} = 6.994)$ prior distribution used in 
Section \ref{sec:model}, we can use a Beta($\tilde \alpha=98.82,\, \tilde \beta 
= 3.74)$ prior on $R/(R + k)$. The posterior distribution of $R/(R + k)$ then 
has a Beta distribution with parameters
\begin{IEEEeqnarray*}{rCl'rCl}
\alpha & = & \tilde \alpha + \sum_{ s = t - \tau + 1}^t I_s, &
\beta & = & \tilde \beta + k \sum_{s = t - \tau}^{t-1} I_s.
\end{IEEEeqnarray*}
While the hyperparameter values are not so small as to be uninformative, they 
are easily outweighed by the data in most settings. By a change of 
variables from $R/(R + k)$ back to $R$, we derive the 
posterior distribution of $R$ to be
\begin{equation}
    \label{eq:R posterior trivial}
    p\left(R \middle| \vec{I}_{[t]},k\right) \propto \frac{k}{(R+k)^2} 
\left(\frac{R}{R + k}\right)^{\alpha - 1} \left( \frac{k}{R + k} \right)^{\beta 
- 1}.
\end{equation}
As a final simplifying step, we compute the normal approximation of this
posterior \cite[Section 4.1]{gelmanBDA04}. To this end, the first and second 
derivatives of the 
$\log$-posterior density are
\begin{IEEEeqnarray*}{rCl}
  \frac{d}{dR} \log p(R|\vec{I}_{[t],k}) & = &\frac{ \alpha - 1}{R} - 
\frac{\alpha + \beta}{R + k},\\  
\frac{d^2}{dR^2} \log p(R|\vec{I}_{[t]},k) 
  & = & -\frac{\alpha - 1}{R^2} + \frac{\alpha + \beta}{(k + R)^2}.
\end{IEEEeqnarray*}
Thus, the mode of the posterior is
\begin{IEEEeqnarray*}{rCl}
\hat R = \frac{k(\alpha - 1)}{\beta + 1}, 
\end{IEEEeqnarray*}
and the variance estimate is
\begin{IEEEeqnarray*}{rCl}
\left( -\frac{d^2}{dR^2} p(R | \vec{I}_{[t]}, k)(\hat R) \right)^{-1} = 
\frac{k^2(\alpha + \beta)(\alpha -1)}{(\beta + 1)^3}. 
\end{IEEEeqnarray*}
This yields a normal approximation of the posterior of
\begin{IEEEeqnarray*}{rCl}
p(R|\vec{I}_{[t]},k) & \approx & N\left(\frac{k(\alpha - 1)}{\beta + 
1},\frac{k^2(\alpha + \beta)(\alpha -1)}{(\beta + 1)^3}  \right). 
\end{IEEEeqnarray*}

\subsection{Generation Model}

It is easiest to represent the process of infections per generation if we 
allow the indices of the summation notation to be real numbers (hence treating 
the summation as integration) via
\begin{IEEEeqnarray*}{rCl}
\sum_{s=c_1}^{c_2} I_{t-s}
  & = & 
  (\lceil c_1\rceil - c_1)I_{t - \lceil c_1 \rceil + 1} + 
  \sum_{s = \lceil c_1 \rceil}^{\lfloor c_2 \rfloor-1} I_{t-s} +
  (c_2 - \lfloor c_2 \rfloor)I_{t-\lfloor c_2 \rfloor}
\end{IEEEeqnarray*}
for $c_1,c_2\in\mathbb{R}$ where $c_1 < c_2$. When $D_{g}$ is the length of 
a generation, the number of infections per generation is then given simply by
\begin{IEEEeqnarray*}{rCl}
  \tilde I_{\tilde t - i} & = & 
  \sum_{s = i\cdot D_{g}}^{(i+1)\cdot D_{g}} I_{t-s},
\end{IEEEeqnarray*}
for $i \in \mathbb{N}_0$.

We assume $R$ is constant for $\tau$ days in the generation model as in the 
momentum model. The corresponding parameter in the generation model is 
$\tau_{g} 
:= \tau/D_{g}$, where we account for non-integer values as before by summing 
fractional daily infections. Estimating parameters and producing forecasts 
requires similar modifications for non-integer values. Conceptually, however, 
these correspond to the same sums as before, just over generations instead of 
conventional days. This can be represented concisely in the notation for 
real-valued summation as
\begin{IEEEeqnarray}{rCl}
\alpha 
  & = & \tilde\alpha + \sum_{s=0}^{\tau_{meta}} \tilde I_{\tilde t-s}, \quad 
\text{and}\\
\beta 
  & = & \tilde \beta + k\sum_{s=1}^{\tau_{meta} + 1} 
  \tilde I_{\tilde t-s}.
\end{IEEEeqnarray}

This yields a negative binomial observation model as before:
\[
\tilde I_{\tilde t} | R, \tilde I_{\tilde t-1} \sim NB\left( \tilde 
I_{\tilde t-1}k, \frac{R}{k+R}\right).
\]

For prediction and comparisons used in Section \ref{sec:austria}, it is more 
sensible to compare the cumulative incidence of several, say $\mathbf T$, days. 
We forecast $\lceil \frac{\mathbf T}{D_{g}} \rceil $ generations $\tilde 
I_{\tilde t}$, and our forecast for $\mathbf T$ days is then
\[
X_{\mathbf T} = \sum_{s = 1}^{\lfloor \frac{\mathbf T}{D_{g}} \rfloor} 
  \tilde I_{\tilde t+s} + 
  \left(\frac{\mathbf T}{D_{g}} -  
  \bigg\lfloor \frac{\mathbf T}{D_{g}} \bigg\rfloor \right) 
  \tilde I_{\tilde t+\lceil \frac{\mathbf T}{D_{g}} 
\rceil}.
\]
In the case study of Section \ref{sec:austria}, we forecast the total weekly 
cases, i.e., $\mathbf T = 7$. Observe that this equates to merely forecasting 
sufficient generations to cover the desired time period, then taking the 
appropriate proportion of the final forecasted generation to match the desired 
time window $\mathbf T$.

\section{Results for Croatia and Czechia}
\label{sec:results2}

\begin{figure}
  \begin{subfigure}{1\textwidth}
   \centerline{\includegraphics[width=1\textwidth, trim=0 0 0 0, 
clip]{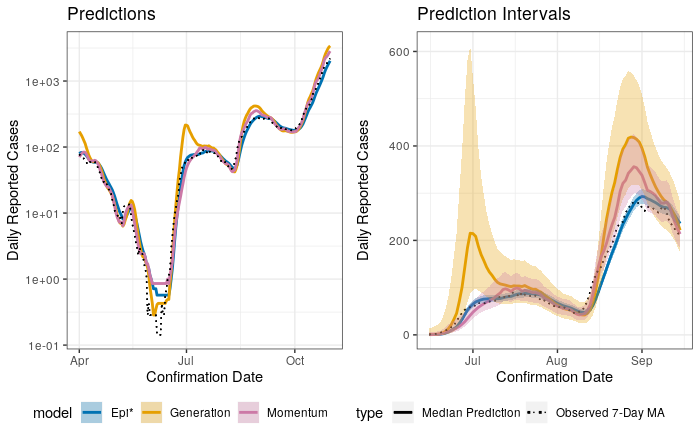}}
   \caption{Croatian Data}
   \label{fig:croatiaPI}
  \end{subfigure}
  \begin{subfigure}{1\textwidth}
   \centerline{\includegraphics[width=1\textwidth, trim=0 0 0 0, 
clip]{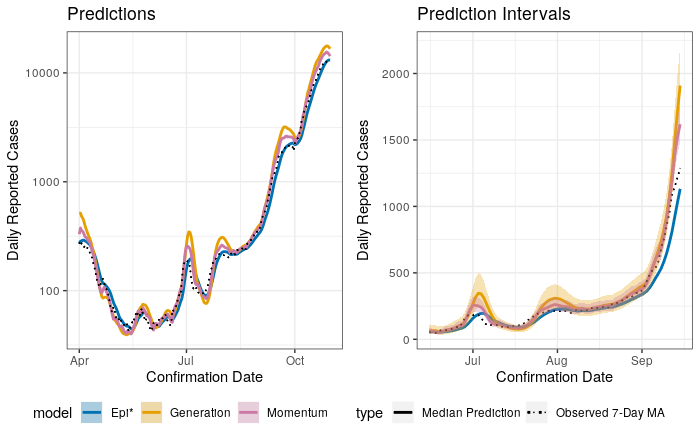}}
   \caption{Czech Data}
   \label{fig:czechiaPI}
  \end{subfigure}
  \caption{Predictions between April 1 and October 31, 2020, and 90\% prediction 
intervals between June 15 and September 7, 2020. Predictions and intervals are 
for the 7-day average of new cases in the following week in Croatia and 
Czechia.}
\label{fig:other-pi}
\end{figure}

\begin{figure}
  \begin{subfigure}{1\textwidth}
   \centerline{\includegraphics[width=1\textwidth, trim=0 0 0 0, 
clip]{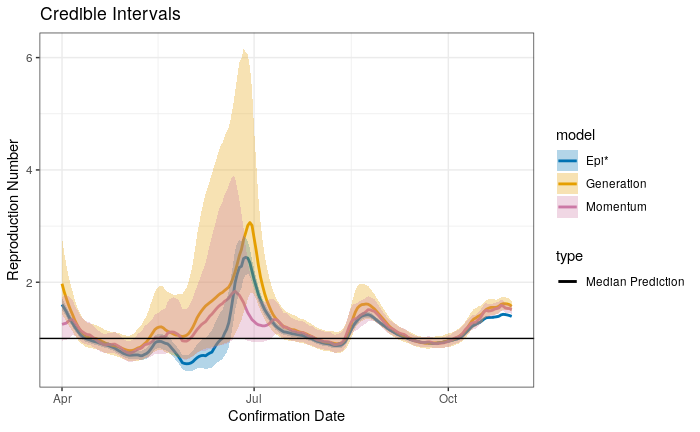}}
   \caption{Croatian Data}
   \label{fig:croatiaCI}
  \end{subfigure}
  \begin{subfigure}{1\textwidth}
   \centerline{\includegraphics[width=1\textwidth, trim=0 0 0 0, 
clip]{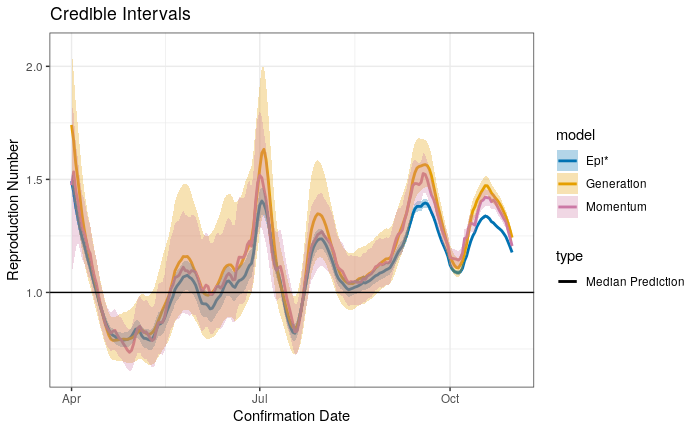}}
   \caption{Czech Data}
   \label{fig:czechiaCI}
  \end{subfigure}
\caption{Credible intervals and for R in Croatia and Czechia between April 1 
and October 31, 2020.}
\label{fig:other-ci}
\end{figure}

As further demonstration of the momentum model, Figures \ref{fig:other-pi} and 
\ref{fig:other-ci} show the same prediction and estimation results as seen in 
Section \ref{sec:austria} but for Croatia and Czechia. The disease 
progression in Czechia is similar to that of Austria over the shown 
period. Croatia is a common Austrian and Czech tourist destination and the 
disease progression is markedly different there than in Austria. The estimated 
coverage probabilities of the prediction intervals are also shown in Table 
\ref{tab:coverage}. The story remains the same as before: coverage is far better 
for the momentum model with superpreading than without (Epi*). Similarly, Epi* 
appears shifted relative to the observed cases, particularly for the Czech data. 
Here we see that the momentum model performs better than the generation model, 
particularly around peaks in the time series.

Figure \ref{fig:other-ci} contains results on the estimation of $R$ for Croatia 
and Czechia. The results are qualitatively the same, in that the 
momentum model with superspreading produces much wider credible intervals. One 
obvious feature of the Croatian data, however, is a steep decline and subsequent 
steep increase in June. This corresponds to a large increase and plateau in 
cases as seen in Figure \ref{fig:other-pi}. The Epi* model estimates that $R$ 
increases to well over 2 within a short period of time before decreasing again 
to previous levels. Alternatively, in the same period, the momentum model 
provides a noticeably lower median estimate but with an incredibly wide 
interval. Further exploration of the feature is warranted, though it is 
reasonable that such a large deviation over a small window of time should 
produce significantly more uncertainty in the value of the underlying parameter, 
particularly when the model is estimated under the assumption that $R$ is 
constant over $\tau=13$ days. Within the momentum model, such short-term 
deviations can be captured by an increase or decrease in disease momentum 
instead of just an increase in $R$. On the other hand, this feature appears to 
show a flaw within the generation model, as both the estimated $R$ and interval 
estimate have extreme spikes. This is likely due to the short-term nature of the 
case increase and the generation model only using roughly three generations for 
estimation.

\end{document}